# Design of Patchy Particles using Ternary Self-Assembled Monolayers

**Inés C. Pons-Siepermann,**[a] **and Sharon C. Glotzer**[a,b]

Recent simulations have studied the formation of patterns in a binary mixture of immiscible surfactants absorbed onto the surface of a spherical nanoparticle. The resulting patterns (Janus, spots and stripes) were in good agreement with experimental results. We perform dissipative particle dynamics (DPD) simulations to study the patterns obtained by adding a third surfactant to the monolayer as a guide towards increasing the richness and diversity of patchy particles synthesized this way. We predict a variety of new patterns that can be produced through different combinations of simple design elements, like nanocolloid size, degree of surfactant immiscibility, stoichiometry of the monolayer, and length difference between surfactants. In all cases, free energy minimization through conformational entropy maximization determines equilibrium pattern formation.

*Submission date: January 3, 2012*

## Introduction

Monolayers of alkanethiols on flat gold surfaces have been widely studied,[1] including the effect of varying the end group[2] and the length of the surfactants.[3] Similar computational studies have also been performed on monolayers of alkanethiols on spherical gold nanoparticles, forming a type of monolayer protected metal nanoparticle (MPMNP).[4] It was shown both through experiments and simulations that if a binary monolayer formed by a mixture of surfactants with different end groups is adsorbed onto a flat gold surface, the surfactants separate into discrete irregular domains on the nanometer scale.[5]

Stellacci and coworkers were the first to extend these studies of binary alkane-thiol monolayers to spherical particles.[6] They showed that when the nanoparticle (NP) is coated with a binary mixture of immiscible surfactants (different end-groups to the alkane chain) of different length, it was possible to obtain either Janus particles or stripes on the surface of the NP. They further showed the possbility to tune the width of the stripes by varying the composition of the binary SAM.[6] Computer simulations showed that the stripes form because of entropic gains due to the length (or tail "bulkiness") mismatch between the surfactants.[7] The free energy of a mixture is normally reduced by decreasing the interface between the immiscible species, which in the case of a binary SAM MPMNP results in the formation of a Janus particle.[7,8] However, longer and bulkier surfactants gain free volume by aligning next to smaller or less bulky ones, thus increasing their conformational entropy. When this entropy gain is sufficient to compensate for the energetic penalty due to the creation of interfaces, the surfactants microphase separate into patchy domains, including stripes and spots (2D micelles).[7,8,9]

There are several promising applications for this type of nano-patterned particle,[10] but constructing diverse structures necessitates richly patterned building blocks. One possible way of producing these richly structured building blocks from MPMNPs is to increase the number of surfactants in the SAM. Analogous generalizations are well-known for block-copolymers, where progressing from diblock to triblock copolymers produces many new phases,[11] some of which are straightforward extensions of diblock copolymer phases, while others have no diblock analogue Although multicomponent SAMS on NPs may be nontrivial to produce experimentally, MPMNPs with up to five different thiol functional groups have been reported in the literature.[12]

Here we use computer simulations to predict the patterns that might be obtained by adding a third surfactant to the SAM that coats the NPs. We first present the results from varying (i) the size of the NP, (ii) degree of immiscibility, (iii) surfactant length, and (iv) SAM stoichiometry, and discuss how each of these parameters modifies the patterns. We then discuss the method and model used in our simulations. We summarize the design rules that result from our simulations in a concluding section.

## Results

### Effect of NP size and degree of immiscibility

The effect of NP size (measured by particle radius) and the degree of immiscibility (measured by the interbead repulsion $a_{ij}$) are presented in Figure 1. Dissimilar surfactants separate without forming stripes for the smallest NP radius studied. When the curvature of the NP provides enough free volume for surfactant tails to explore, the possible entropic gains from stripes are not enough to overcome the energetic penalties that accompany interfaces. The analogous effect is observed for binary systems, where a Janus particle forms for smaller NP radius.[7,8] As seen in Figure 1, for ternary systems a Cerberus (or "three-faced") particle is formed when the three surfactants adopt the minimum interface possible between them.

As NP radius increases, the surfactants assemble into striped patterns. For the conditions of Figure 1, the short surfactant (red in the Figures) is always observed between the long (yellow) and medium (blue) ones. This pattern apparently minimizes free energy by providing enough length difference to support the formation of the stripes, thereby maximizing conformational entropy as in the binary systems of reference 7.

For the smallest interbead repulsion studied ($a_{ij}$ = 30), a pattern with Alternating Stripes is formed. In this pattern, the medium and long surfactants form alternating stripes with the short ligand always between them. As the NP radius increases, the stripes become disordered as observed in Figure 1 for a NP radius = 8. The effect of disordered stripes for large NP radius was reported also in binary systems.[7]

From the Alternating Stripes pattern it is evident that the entropic gains are only sufficient to form stripes when either the long or medium ligands form stripes with the short surfactant. However, the entropy gain from stripes formed between the medium and the long ligands are not sufficient to offset the enthalpic penalties of creating interfaces and therefore these types of stripes are not seen. Since the length difference between the short and the medium ligand, and between the medium and the long ligand, are the same (3 beads) (and therefore the gain in free volume is the same), but stripes are only observed for the first case, it is evident that the relative length of the surfactants with respect to the NP radius is also an important factor in the formation of stripes. This effect will be discussed in the next section.

For higher interbead repulsion ($a_{ij}$ > 30), different striped patterns appear. The Cerberus particle transitions into a Neapolitan particle with three parallel stripes. In this pattern the long and medium surfactants occupy the two poles of the NP, while the short surfactant forms a single stripe between them. As the NP radius increases, the short ligand mixes with the long ligand to form stripes on one face of the NP while the medium length surfactant remains separated on the other face, forming a Striped Janus particle. We find the long and short surfactants behave as in a binary system, with increasing number and disorder of stripes as the NP radius increases.[7]

Increasing the interbead repulsion increases the energetic penalties for forming stripes. A smaller radius of curvature (i.e. less available free volume for the tails) is thus needed before the striped patterns are favoured. This means that as the ligand immiscibility increases, the transition from Cerberus to Neapolitan to Striped Janus requires a larger increase in the NP radius.

**Effect of surfactant length**

Figure 2 shows the effect of surfactant length on the formation of patterns. We investigated the effect of changing one, two, or all three ligands, resulting in a three-dimensional design space where the length of each surfactant corresponds to each dimension. In Figure 2a, a view of the effect of the length of the short and medium surfactants is shown, and the axis corresponding to the long surfactant is not presented. For surfactants relatively short compared to the NP radius, we observe a Striped Janus particle. As the length of the surfactants increases with respect to the NP radius, the system evolves to Neapolitan and Cerberus particles. The stripes are lost because the length difference between the surfactants does not provide enough entropic gain to overcome the high energetic penalties due to having long chains of unlike beads forming interfaces. This is the same effect observed in Figure 1, where given the same length difference between two surfactants, we find only the smaller ones (short, 3, and medium, 6 beads) form stripes, whereas the longer ones (medium, 6, and long, 9 beads) do not.

Figure 2b through c show planes perpendicular to the one shown in Figure 2a, corresponding to the red rectangles. In Figure 2b and c, the patterns obtained as a function of length of the medium and long surfactants are shown, keeping the length of the short surfactant constant in each case. Three different cases were studied for each plane, corresponding to the numbered lines in the figures.

The first case, Line 1, is parallel to the long surfactant axis. The length of the medium surfactant is kept constant at just one bead longer than the short surfactant, as the length of the long surfactant is increased to up to eight beads longer than the short one. The net effect of this case is a SAM formed by two short and one long surfactant. For short surfactants (Figure 2b) we find that a one-bead difference is still sufficient to stabilize the formation of stripes. However, as the length of the surfactants increases (c and d), we observe Cerberus particles. In this last case, for a NP of radius 4, surfactants of 10, 11, and 12 beads act essentially as same-length surfactants and phase separate to minimize the interface between them. Moving from left to right along Line 1, Striped Janus particles begin to appear as the long surfactant grows. However, unlike the cases previously studied (Figure 3a), now an interface between the long and medium surfactant forms, without the need for the short ligand between them (Figure 3b). This pattern persists as the medium surfactant becomes sufficiently short with respect to the long surfactant to offer enough gain in the free volume to justify the formation of stripes.

The second case, Line 2, corresponds to the same length difference between the short and medium as between the medium and long surfactants. We observe Alternating Stripes when the surfactants are short relative to the NP radius (Figure 2b), but as they grow in length this pattern is lost, becoming first Striped Janus (Figure 2c) and finally Neapolitan particles (Figure 2d).

The last case, Line 3, shows the case when the length difference between the long and medium surfactant is only one bead, essentially acting as two long surfactants and one short. For this case we observe behavior similar to that of Line 1. However, in this case the short surfactant always appears between the other two. Also, we find that the Janus particle exhibits stripes on both sides, instead of only one (Figure 3c). Since the length difference between the long and medium surfactant is only one bead, the entropic gains from forming stripes with either one of them are apparently similar so stripes form on both sides. Again this effect is lost as the surfactants become too long with respect to the NP radius, with the number of stripes decreasing in (Figure 2c), and a Neapolitan particle forming in (Figure 2d).

**Effect of SAM stoichiometry**

For fixed NP radius = 4 and length of surfactants (3, 6, and 9 beads), the effect of SAM stoichiometry is presented in Figure 4. For interbead repulsion = 30 (Figure 4a), we observe the Alternating Stripes pattern around the 1:1:1 composition of the SAM. When the concentration of one surfactant is >70%, the patterns take the form of 2D micelles – spots – instead of stripes (Figure 5). When either the medium or long surfactants are in the highest concentration, we observe a Janus particle with micellar patches of the short surfactant dispersed into the surfactant that is present in the highest concentration (Figure 5a and b). When the short surfactant is in the highest proportion, we find the other two form micelles dispersed in a continuous matrix of the short surfactant (Figure 5c). In this case, the entropic gains achieved by the creation of additional interfaces with the short surfactant are apparently sufficient to stabilize this pattern. For the cases when there are only two surfactants present in the SAM, we find stripes also for stoichiometries where the relative concentration of the surfactants is similar. If one of them is in excess, we observe micellar patches except for the case where there is no short surfactant present. In this case, we observe Janus particles, and even for near 1:1 stoichiometries the stripes formed are much thicker than the ones formed in all other cases. This is due to the effects of surfactant length as previously discussed.

For the higher interbead repulsion (Figure 4b), we always observe the Striped Janus pattern. For very low concentrations of the small surfactant (10%) we observe a Neapolitan particle with a very thin stripe of the small surfactant separating the other two. When there is no small surfactant present, the other two separate to form a Janus particle with no stripes. In this case, the relative length of the two surfactants with respect to the NP radius becomes too high, and no stripes are observed.

## Method and Model

We use dissipative particle dynamics (DPD),[13] a coarse grained algorithm that simulates the behaviour of fluids using interacting beads that move in a continuous space during discrete time steps in the NVT ensemble. Mass and momentum are conserved throughout the system, which is in thermal equilibrium and satisfies the fluctuation-dissipation theorem.[14] The repulsive nature of the force between beads favours the modelling of phase separation of immiscible surfactants in which the net effective interaction between unlike surfactants is repulsive. DPD has been used successfully to model block copolymers,[15] amphiphilic mesophases,[16] surfactant[17] and polymer phase separations,[18] as well as the assembly of patchy patterns on binary MPMNs.[7]

The surfactants are represented as chains of beads connected by simple harmonic springs. The thiol heads are directly adjacent to the surface of the NP and restricted to the surface using constrained dynamics.[19] The diameter of the beads, $\sigma$, is set to unity; all lengths are presented in units of $\sigma$, except for ligand length which is expressed as numbers of beads. We measure mass in units of the bead mass $m$, and energy in units of $k_BT$, where $m$ and $k_BT$ are also unity ($k_B$ is the Boltzmann factor and T the temperature). Interbead repulsion between beads of the same surfactant is $a_{ij} = 15$, and for different surfactants is varied from 30 (weakest immiscibility) to 365 (strongest immiscibility).

We simulated both binary and ternary SAMS with a surface density of 6 beads/$\sigma^2$. We used a time step of 0.02 for all runs and confirmed using a time step 1/10 as large that the results are converged and independent of discretization time step. To ascertain that the patterns presented correspond to equilibrated states, we simulated different initial conditions (including mixed random, phase separated, and striped medium-short-long, medium-long-short, and long-medium-short) to confirm that the resulting patterns are independent of thermodynamic path.

## Conclusions

The formation of patterns on NP coated with ternary SAMs was studied. New patterns were predicted, including Cerberus, Neapolitan, Striped Janus, Alternating Stripes, and Spotted particles. As for binary SAMs,[7] in all cases, the patterns formed can be explained by free energy minimization. When the difference in lengths leads to a sufficient conformational entropy gain at interfaces, then the interface will increase to maximize free volume accessible to the ligand tails. When the free volume gain across interfaces is small, then the particles will become increasingly segregated, changing from Spots to Alternating Stripes to Striped Janus to Neapolitan and finally Cerberus particles.

All of these patterns are obtainable by manipulation of simple design elements, such as NP radius, degree of immiscibility between surfactants, stoichiometry of SAM, and length of surfactants. We can summarize the design rules we predict here as follows:
- Small NP radius produces a Cerberus pattern;
- Small length differences between even weakly immiscible surfactants also produces Cerberus patterns;
- Symmetric length differences between surfactants produces an Alternating Stripes pattern for weak ligand immiscibility;
- Symmetric length differences between surfactants produces a Striped Janus particles for strong ligand immiscibility;
- Increasing the NP radius for Striped particles induces disorder in the stripe pattern;
- One short and two long surfactants produces a Striped Janus particle but alters the ordering of the stripes;
- Two long and one short surfactant produces a Striped Janus particles with stripes on both sides of the Janus particle;
- Increasing the concentration of one of the surfactants induces the formation of a spotted particle covered by 2D micelles.

## Acknowledgements

This material is based upon work supported by the Defense Threat Reduction Agency under Grant No. HDTRAI-09-1-0012. SCG is also supported by a National Security Science and Engineering Faculty Fellowship from the DOD/DDRE under Grant No. N00244-09-1-0062. Any opinions, findings, and conclusions or recommendations expressed in this material are those of the author(s) and do not necessarily reflect the views of the Defense Threat Reduction Agency or those of the DOD/DDRE. ICPS and SCG also acknowledge support from the James S. McDonnell Foundation 21st Century Science Research Award/Studying Complex Systems, Grant No. 220020139.

## Bibliographic references and notes

*[a] Department of Chemical Engineering, University of Michigan. 3440 G.G.Brown Building, 2300 Hayward St., Ann Arbor, Michigan, US. Tel: 734 763 9895; E-mail: inespons@umich.edu*

*[b] Department of Materials Science and Engineering, University of Michigan. 3406 G.G.Brown Building, 2300 Hayward St., Ann Arbor, Michigan, US. Fax: 734 764 7453; Tel: 734 615 6296; E-mail: sglotzer@umich.edu*

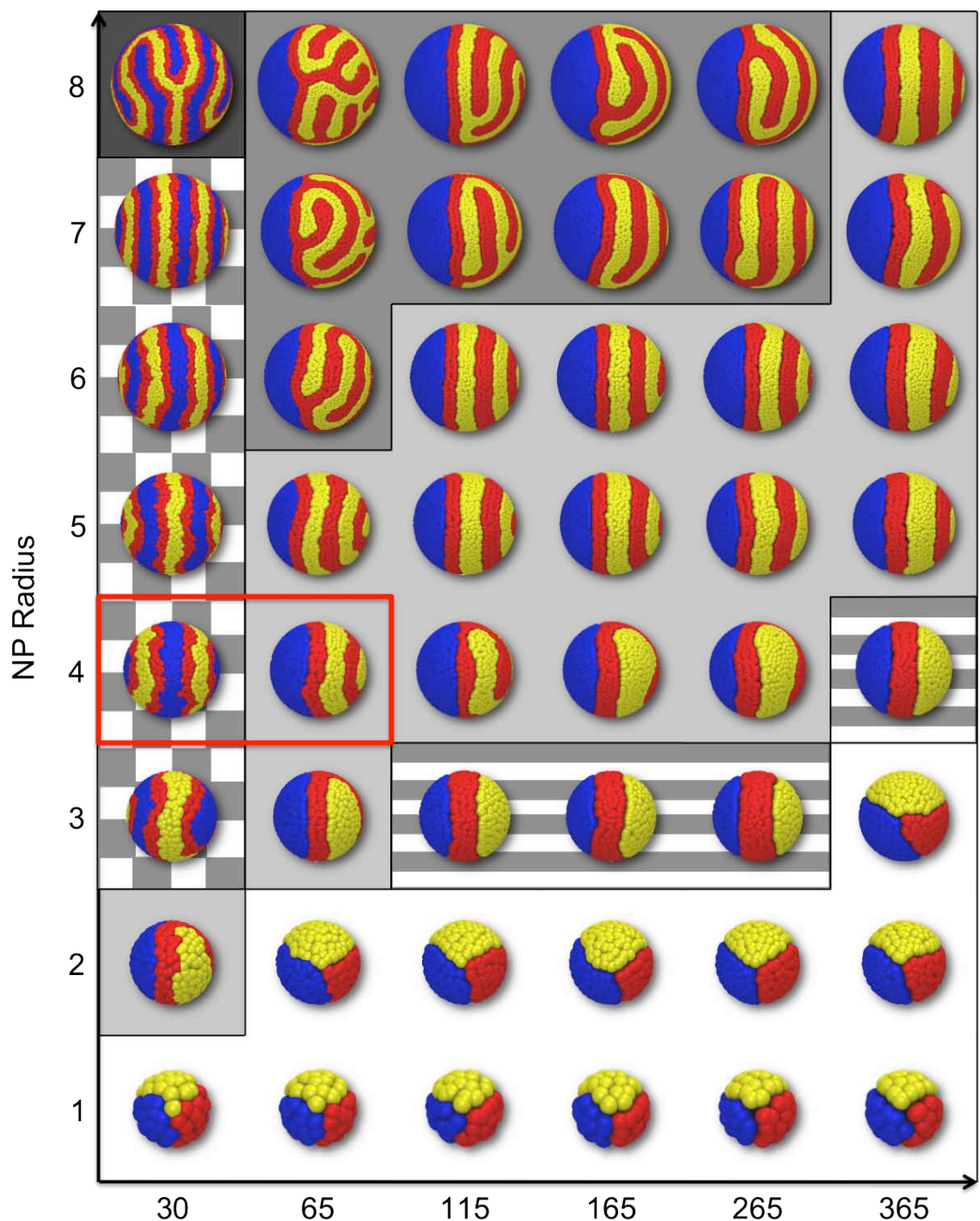

Figure 1 Effect of degree of surfactant immiscibility as controlled by interbead repulsion vs. NP radius. Tails not shown. Length of surfactants is 3 (red), 6 (blue), and 9 (yellow) beads. Composition of SAM is 1:1:1. NPs not drawn to scale. Radius of NPs in top row is eight times that of bottom row. Interbead repulsion increases from 30 to 365 from left to right. Similar background pattern used to help identify similar "phase" for different parameter combinations. For systems highlighted by red rectangle, effect of surfactant length SAM stoichiometry has been shown in Figure 2 and Figure 4.

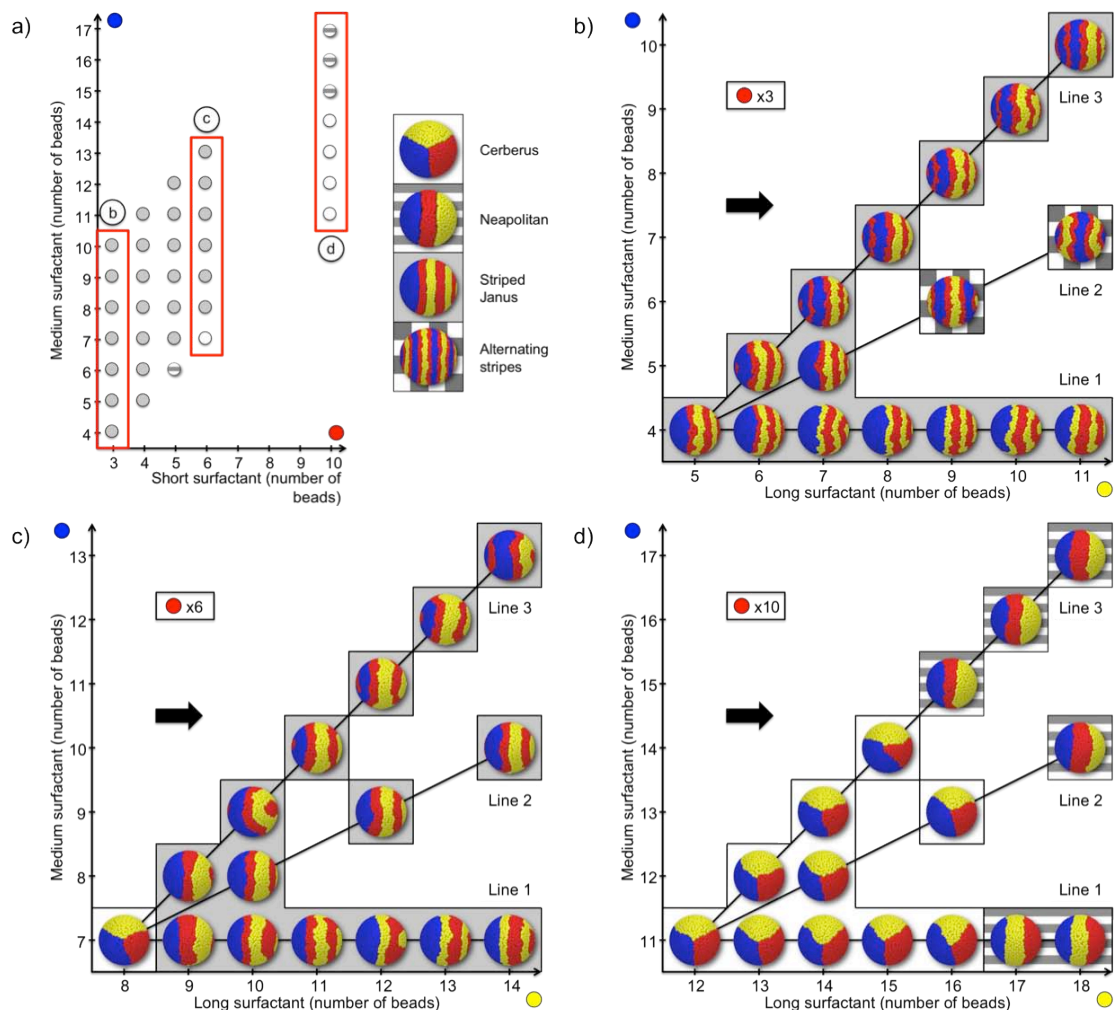

Figure 2 Effect of surfactant length for low surfactant immiscibility. a) Length of short surfactant vs. length on medium surfactant. Dots represent data points simulated for different surfactant lengths and the patterns obtained are presented according to the coloring of Figure 1. b) through d) Length of medium surfactants vs. length of long surfactant. b) through d) are side views of planes highlighted by red squares in a), after adding a third axis representing the length of the long surfactant. NPs have radius 4. Stoichiometry of SAM is 1:1:1. Interbead repulsion between unlike surfactants is 30. Length of the short surfactant is b) 3, c) 6, and d) 10 beads. Length of medium and long surfactants varies along the axis in the figures. The black arrows represent the side from which the figures are viewed in a). In b) through d), line 1 corresponds to the case with two short surfactants and one long, with only a one-bead difference between the short and medium surfactant. Line 2 represents the case when the difference in number of beads between the short and medium surfactant is equal to the difference in number of beads between the medium and the long. Line 3 represents the case when there are two long surfactants in the system, with the difference between the medium and the long being only one bead. Tails not shown.

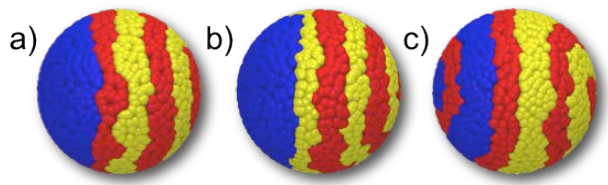

Figure 3 Striped Janus particles. a) Short (red) and long (yellow) surfactants form stripes, while medium ligand (blue) separates. No interface between medium and long. b) Same as a), but with interface between medium and long surfactants. c) Short surfactant forms stripes on both sides of Janus particle. Tails not shown.

a)

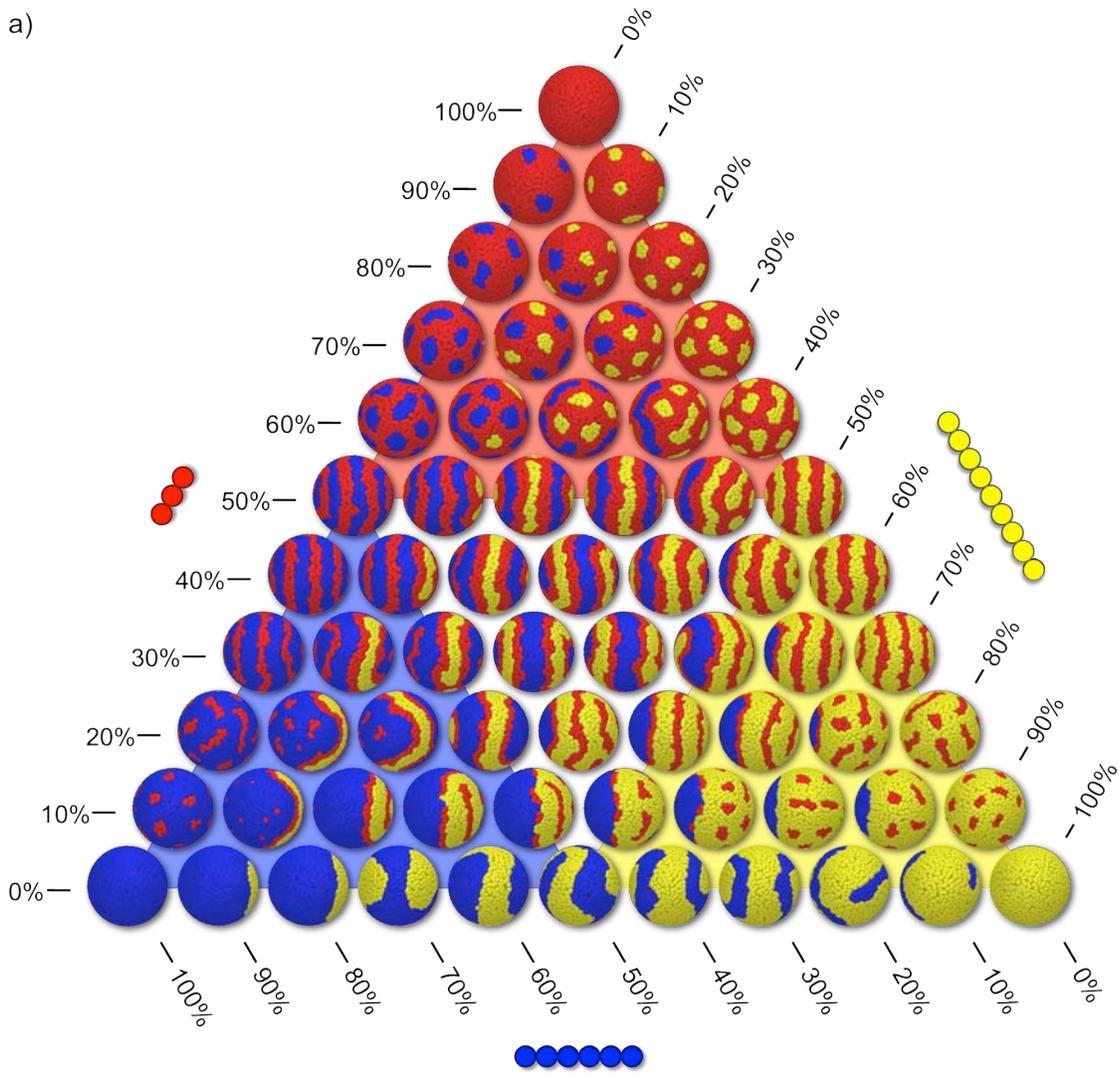

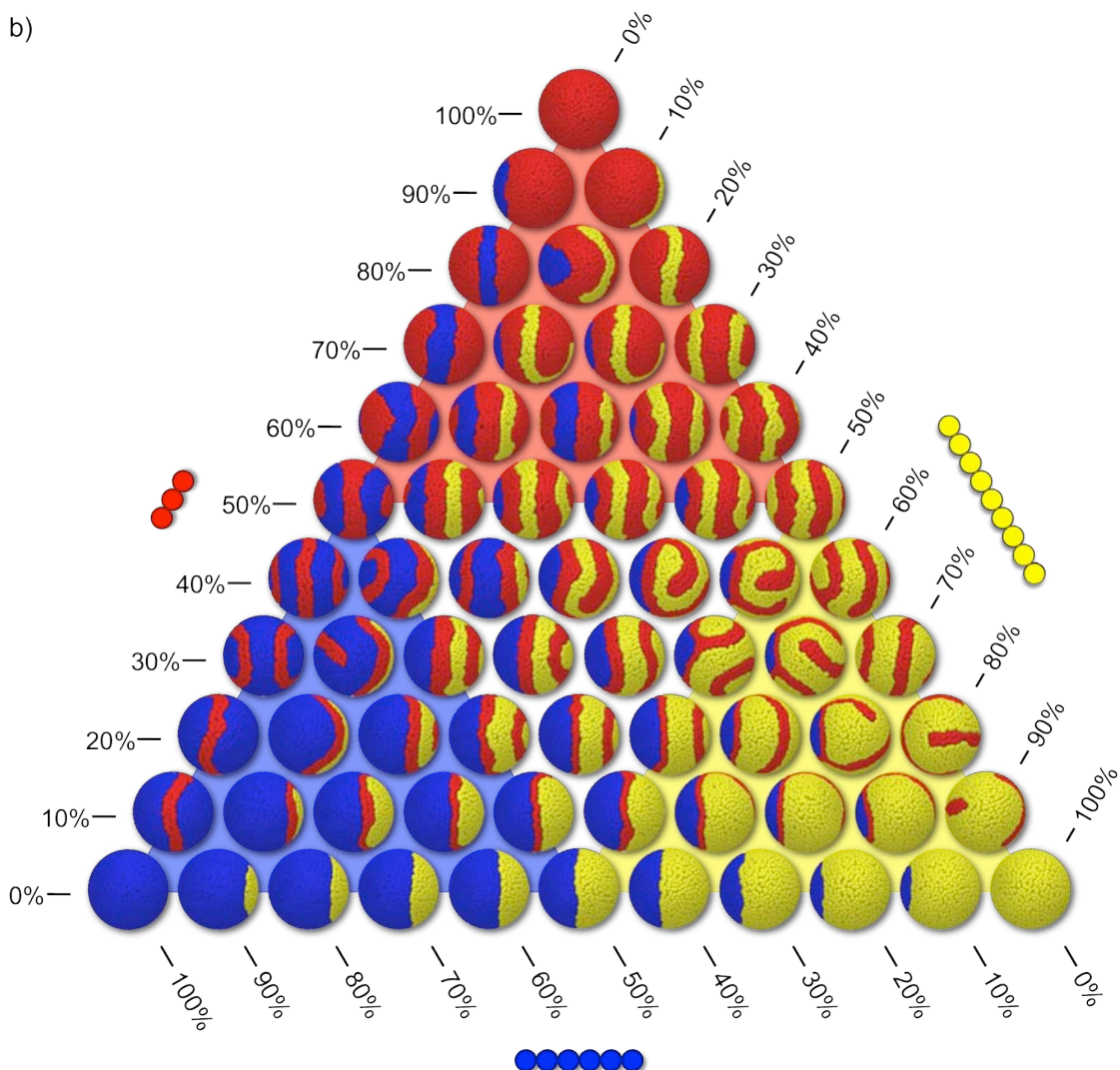

Figure 4 Effect of SAM stoichiometry. Only head beads shown. NPs have radius 4, and length of surfactants is 3, 6, and 9 beads. Interbead repulsion is a) 30 (weak immiscibility) and b) 65 (strong immiscibility). Color shadows in background represent the surfactant in highest concentration in SAM. Tails not shown.

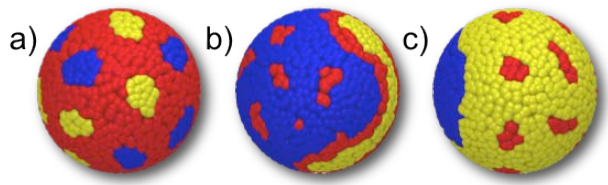

Figure 5 Spotted patterns of 2D-micelles, obtained for concentrations >70% of a) short, b) medium, and c) long surfactant. Tails not shown.